# Knowledge-Based Three-Dimensional Dose Prediction for Tandem-And-Ovoid Brachytherapy

Running title: 3D Dose Prediction for GYN Brachytherapy

Katherina G. Cortes, Aaron Simon, Karoline Kallis, Jyoti Mayadev, Sandra Meyers, and Kevin L. Moore*

Department of Radiation Medicine and Applied Sciences

University of California San Diego

La Jolla, CA 92093

*Corresponding author: kevinmoore@health.ucsd.edu

## Abstract

**Purpose:** The purpose of this work was to develop a knowledge-based voxel-wise dose prediction system using a convolution neural network (CNN) for high-dose-rate brachytherapy cervical cancer treatments with a tandem-and-ovoid applicator.

**Methods:** A 3D U-NET CNN was utilized to output voxel-wise dose predictions based on organ-at-risk (OAR), high-risk clinical target volume (HRCTV), and possible source location geometry. The available dataset was a five-year sample of previously treated tandem-and-ovoid treatments comprising 397 cases (273 training: 61 validation: 61 test). HRCTV, OARs (bladder/rectum/sigmoid). Structures and dose were interpolated to 1mmx1mmx2.5mm dose planes with 2 channels: one for dose-emitting structures (possible source positions) and the other for voxel identification (OAR, HRCTV, or unspecified) with a single output channel for dose. To assess 3D voxel prediction accuracy, we evaluated the dose difference $\delta D_{xyz,ij} = D_{actual,ij}(x,y,z) - D_{predicted,ij}(x,y,z)$ and dice similarity coefficients in all cohorts across the clinically-relevant dose range for cervical cancer brachytherapy (20-130% of prescription), mean and standard deviation. We also examined discrete DVH metrics utilized for tandem-and-ovoid plan quality assessment: HRCTV D90%(dose to hottest 90% volume) and bladder/rectum/sigmoid D2cc, with $\Delta D_x = D_{x,actual} - D_{x,predicted}$ Pearson correlation coefficient, standard deviation, and mean quantifying model performance on the clinical metrics.

**Results:** Voxel-wise dose difference accuracy for 20-130% dose range inside contours volumes for training (test) ranges for mean $\overline{\delta D}$ and standard deviation $\sigma$ for all voxels [-0.3%±2.0% to +1.0%±12.0%] ([-0.1%±4.0% to +4.0%±26.0%]), HRCTV [-3.5%±5.1% to -1.7%±12.8%] ([-3.5%±4.8% to -2.6%±18.9%]), bladder [-0.7%±2.4% to +3.2%±12.0%] ([-2.5%±3.6% to +0.8%±12.7%]), rectum [-0.7%±2.4% to +15.5%±11.0%] ([-0.9%±3.2% to +27.8%±11.6%]), and sigmoid [-0.7%±2.3% to +10.7%±15.0%] ([-0.4%±3.0% to +18.4%±11.4%]). Voxel-wise dice similarity coefficients for 20-130% dose ranged from [0.96, 0.91] for training and [0.94, 0.87] for test cohort. Relative DVH metric prediction in the training (test) set were HRCTV $\overline{\Delta D}_{90} \pm \sigma_{\Delta D}$=-0.19±0.55 Gy (-0.09±0.67 Gy), bladder $\overline{\Delta D}_{2cc} \pm \sigma_{\Delta D}$ = -0.06±0.54 Gy ($-0.17$±0.67 Gy), rectum $\overline{\Delta D}_{2cc} \pm \sigma_{\Delta D}$ = -0.03±0.36 Gy (-0.04±0.46 Gy) , and sigmoid $\overline{\Delta D}_{2cc} \pm \sigma_{\Delta D}$= -

0.01±0.34 Gy (0.00±0.44 Gy).

**Conclusion:** 3D knowledge-based dose predictions for brachytherapy provide accurate voxel-level and DVH metric estimates that could be used for treatment plan quality control and potentially for fully automated plan generation.

## Introduction

Brachytherapy (BT) is considered the standard of care for definitive radiation treatment for locally advanced cervical cancer that often utilizes an intracavitary device, such as a tandem and ovoids (T&O)[1]. The dose is customized for a patient's specific anatomy to ensure adequate dose to the high-risk cancer tumor volume (HRCTV) while sparing the organs-at-risk (OARs). High quality brachytherapy is an essential component of treating cervical cancer, linked to improved pelvic control and disease-free survival[2,3], but can be challenging and highly labor intensive[4,5]. Recent advancements in 3D imaging and treatment planning for intact cervix BT have permitted the development of new techniques for shaping the dose distribution to permit better sparing of OARs and better dose escalation of the target[6,7]. These techniques, which include increased utilization of interstitial applicators[8–10] in addition to standard intracavitary applicators, however, increase not only the technical difficulty of BT implantation but also brachytherapy treatment planning resource requirements.

Intensity Modulated Radiation Therapy (IMRT) has traditionally presented a similarly labor-intensive and practitioner-dependent planning challenge, where inter-patient anatomical variations coupled with subjectivity of human planners leads to widely documented plan quality variations[5] that can put patients at risk for increased complication[11]. Knowledge-based planning, a method that utilizes inferred correlations between patient anatomical variations and final plan dosimetry[12–14] has been shown to reduce these plan variations[15,16]. It is unknown how variable treatment plans are in the context of BT and use of knowledge-based dose prediction techniques for locally advanced cervical cancer BT has been limited[17]. Published knowledge-based models for gynecologic BT are based on OAR dose-volume histogram (DVH) estimates, whereas current development in external beam knowledge-based models focuses on three-dimensional dose distribution prediction[18–21]. The development of a three-dimensional estimation of expected BT dose distributions could highlight inconsistencies in present planning practices and standardize treatment planning for a standard of care therapy in cervical cancer.

In this manuscript we describe a method for three-dimensional knowledge-based dose distribution prediction in the context of T&O BT using a convolutional neural network. The resultant system uses spatial information about OARs, high-risk clinical target volume (HRCTV), and applicator geometry to predict at a voxel-by-voxel level what the expected dose should be for a high-quality T&O treatment. We then sought to quantify the model accuracy across the dose distribution as well as in clinically-important BT quality metrics. In addition to serving as an objective measure quantifying BT plan quality, applications of three-dimensional dose distribution prediction in BT span the range from real-time quality control tool to the basis for fully automated knowledge-based BT treatment planning.

## Materials and Methods

### Clinical datasets

A total of 395 treatment plans from 126 patients with cervical cancer over a 6-year period (2012-2018, UCSD IRB Project #181609) treated with CT-guided T&O BT fractions were available for this analysis. Single fraction prescription doses ranged from 5.5-8 Gy. The available set of patients and plans was divided into training, validation, and test sets with an approximately 70:15:15 ratio: 273 plans in the training set, and 61 plans each for the validation and test sets. No individual patient's BT fractions were put into more than one set. Table 1 summarizes the tumor stage and fraction number of the training, validation, and test sets. During the time range of this retrospective study there were five radiation oncologists treating cervical cancer at our institution; the patients' respective attending physicians are also shown in Table 1.

| Stage | **Training** $N_{training}$=273 plans | | **Validation** $N_{validation}$=61 plans | | **Test** $N_{test}$=61 plans | |
|---|---|---|---|---|---|---|
| T1 | 90 | 33% | 18 | 30% | 16 | 26% |
| T2 | 135 | 49% | 28 | 46% | 28 | 46% |
| T3 | 47 | 17% | 15 | 25% | 14 | 23% |
| T4 | 1 | 0% | 0 | 0% | 3 | 5% |
| **Physicians** | | | | | | |
| A | 122 | 45% | 28 | 46% | 20 | 33% |
| B | 55 | 20% | 9 | 15% | 14 | 23% |
| C | 73 | 27% | 8 | 13% | 26 | 43% |
| D | 12 | 4% | 4 | 7% | 1 | 2% |
| E | 11 | 4% | 12 | 20% | | |
| **Fraction #** | | | | | | |
| 1 | 65 | 24% | 16 | 26% | 12 | 20% |
| 2 | 63 | 23% | 15 | 25% | 13 | 21% |
| 3 | 60 | 22% | 13 | 21% | 13 | 21% |
| 4 | 53 | 19% | 12 | 20% | 14 | 23% |
| 5 | 32 | 12% | 5 | 8% | 9 | 15% |

*Table 1.* *Breakdown of training, test, and validation datasets by different features such as physicians, fraction number, and stage of each case.*

## Image Processing

All patients were planned in the BrachyVision treatment planning systems (Varian Medical Systems, Inc., Palo Alto, CA). All planning CT scans were at 2.5mm slice thickness with transverse pixel size ranging from 0.893 to 1.367mm. Dose grids were calculated at 1.0x1.0x2.5mm$^3$ resolution. The associated DICOM-RT (CT, RT-STRUCT, RT-DOSE) for each plan was exported from BrachyVision and imported into MATLAB (Mathworks Inc., Natick, MA) using CERR[22]. In MATLAB, the contours in the RT-STRUCT object were converted to image masks. To standardize the image matrices, the CT intensity values, and structure masks were interpolated to the same resolution as the dose grid: $1.0x1.0x2.5mm^3$ voxels.

The brachytherapy applicator is extremely critical to the predictions and dose distributions as it dictates possible source positions. The applicators used were the Varian Titanium Fletcher-Suit-Delclos tandem and ovoid set (1AL07522000 Titanium intrauterine tandem angle 15°, 1AL07522001 Titanium intrauterine tandem angle 30°, or 1AL07522002 Titanium intrauterine tandem angle 45°). Because of the clear visibility of the high-density applicator in CT imaging, image features alone can be used without need to utilize the RT-PLAN object that contained the digitized source positions. Digitized source positions were not used because the aim was to utilize only image-based features without requiring applicator digitization. Using a Density- Based Spatial Clustering of Applications with Noise (DBSCAN)[23] algorithm, a minimum HU threshold of 3000 was input to the clustering algorithm to identify the titanium applicator elements: one tandem and two ovoids. DBSCAN model parameters were tuned to maximize applicator identification accuracy. Best performance for identifying titanium applicators was obtained with DBSCAN parameters epsilon = 3.5 and minimum number of points = 30. The tandem was determined from these three clusters according to which cluster of points included the HRCTV center of mass. The other two structures were then labeled as ovoids.

To pre-condition the planning data for neural network learning, we developed two input channels (applicator mask and anatomical mask) and one output channel (dose matrix). The dimensions of the voxel matrices were bounded by a $10x10x10cm^3$ volume centered on the HRCTV center of mass to ensure a similar field of view, a uniform matrix size for each plan, and computationally efficient neural network training on a clinically-relevant region of interest. At this point each plan was represented by a 3D matrix (100x100x40 voxels) with three channels. Each structure in the anatomical input channel was represented by a different number (1 for HRCTV, 2 for bladder, 3 for rectum, 4 for sigmoid); the applicator channel was one channel with a different number for each ovoid and the tandem. The output channel is composed of the continuous dose values, normalized to the prescription dose to ensure prediction similarity between cases with differing prescription values.

**Dose Transformation**

Unlike EBRT, brachytherapy dose distributions diverge rapidly near the applicator due to the inverse square law. Prediction errors in this high gradient, high dose region can contribute disproportionately to a typical loss function of a neural network (e.g. mean squared error). Notably, the prediction accuracy in these extreme regions of the dose distribution do not contribute to standard clinical quality metrics of BT which are evaluated at the boundary of the HRCTV and in the regions of the OARs. To prevent the voxels very near the applicator from dominating the loss function and degrading accuracy in regions with clinically-relevant dose values, we employed a dose transformation function before inputting the dose channel to the neural network. An arbitrary dose transformation can be represented by an invertible monotonic function $f$ whereby a scaled distribution  The value of this approach is that this invertible function can be easily adjusted to tune the network performance without changing the functional operation of the network. For this work we found that scaling the dose over 150% with a square root function provided unbiased predictions in the dose region below 150%:

$$D' = f(D) = \begin{cases} D \leq 150\%, \ D \\ D > 150\%, \ 150\% + (D - 150\%)^{1/2} \end{cases} \quad (1)$$

trivially inverted by:

$$f(D') = \begin{cases} D' \leq 150\%, \ D' \\ D' > 150\%, \ 150\% + \ (D' - 150\%)^2 \end{cases} \qquad (2)$$

**Neural Network Modeling**

As brachytherapy dose distributions have large dose gradients in all directions and voxel dose can be strongly affected by applicator positions that are offset craniocaudially from the voxel (e.g. voxel position is just superior to ovoid, dose will be dominated by a feature not contained in the same z-plane), we employed a convolutional neural network known as a 3D UNET[24,25]. The neural network was trained on the COMET GPU at the San Diego Supercomputing Center[26] using 16GB NIVIDIA P100 GPU Nodes, 16 input filters, Leaky ReLU activation function, dropout layer with parameter 0.3, initial kernel size of 3, Mean Squared Error (MSE) loss function, and ADAM optimizer to give meaningful predictions.

The model was trained in 20 epoch batches until validation loss stopped decreasing to prevent overfitting. This resulted in a total of 100 epochs. The model at each epoch was saved for later access. The final model was then chosen out of the previous models based on when the validation set loss stopped decreasing.

Figure 1 depicts the entire image processing and neural network training pipeline.

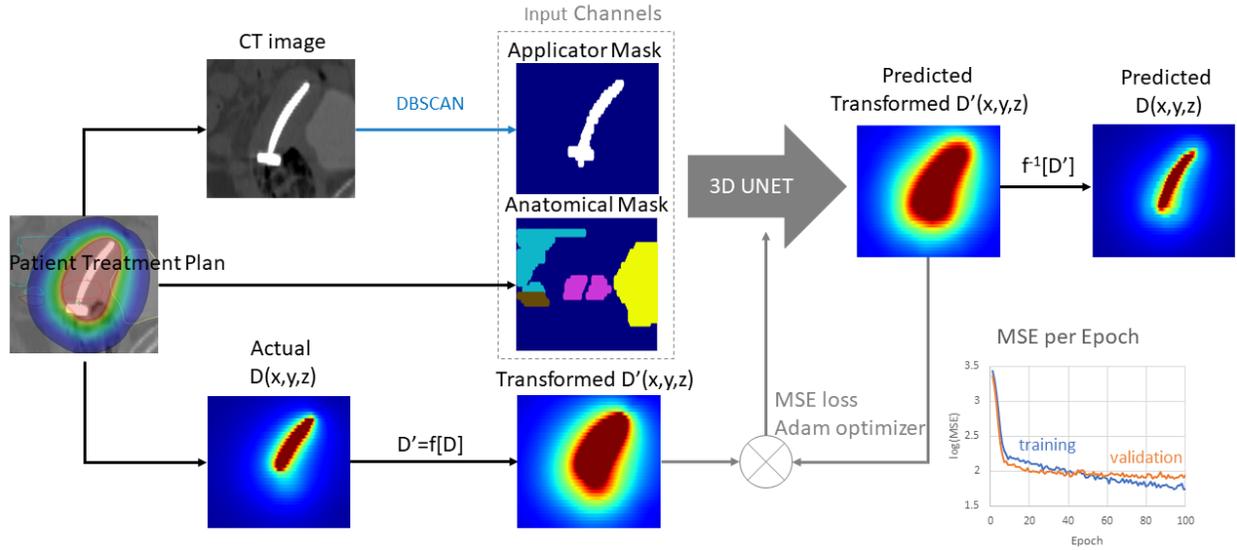

***Fig. 1.*** *Flowchart detailing image processing, neural network training and dose prediction workflow.*

**Evaluating Model Performance**

Model performance was evaluated in two ways: (I) quantifying the voxel-level accuracy of the 3D network predictions and (II) evaluating how well the 3D network predicted clinically relevant quality metrics for gynecologic brachytherapy.

To evaluate the voxel-wise performance, the primary quantification used was the dose difference:

$$\delta D_{xyz,ij} = D_{actual,ij}(x,y,z) - D_{predicted,ij}(x,y,z) \qquad (3)$$

where *x, y, z* represents the Cartesian position of the voxel in the $i$<sup>th</sup> case in the $j$<sup>th</sup> structure, expressed as a percentage of the prescription dose for the case.

Given a dose range $D_l$ to $D_u$, let $S_C(x, y, z, i, j; D_l, D_u)$ represent the set of all points $x, y, z$ in the $i^{th}$ case and $j^{th}$ structure such that $D_l < D_{actual,ij}(x, y, z) \leq D_u$. Structures were defined as HRCTV, bladder, rectum, sigmoid and all voxels irrespective of structure type.

To visualize the prediction accuracy across the C cohorts – training, validation, and test – we examined the voxel-wise dose difference across different percentage deciles using the average dose difference:

$$\overline{\delta D}_{j,C}[D_l, D_u] = \frac{1}{\# \, of \, voxels} \sum_{x,y,z,i \in S_C} \delta D_{xyz,ij} \qquad (4)$$

and its standard deviation:

$$\sigma_{j,C}[D_l, D_u] = \sqrt{\frac{\sum_{x,y,z,i \in S_C}((\delta D_{xyz,ij} - \overline{\delta D}_j[D_l, D_u])^2}{\# of voxels - 1}} \qquad (5)$$

Notably, because these metrics do not include spatial uncertainty terms, it is expected that $\sigma$ will generally increase with dose. Because of this, we evaluated $\overline{\delta D}$ and $\sigma$ for the clinical range 20% - 130% at 10% intervals to understand the dose estimation accuracy (bias and uncertainty) across individual structures, cohorts C, and in different ranges of the dose distribution.

Another metric that has been used[20,27] to quantitatively evaluate 3D dose estimations is the Dice Similarity Coefficient (DSC) computed on isodose surfaces. The DSC is defined as $DSC = \frac{2(A \cap B)}{A+B}$ with output values ranging from 0 to 1, where 1 is perfect prediction of the particular isodose volume. For the purposes of this study, $A_i(D)$ = set of all voxels in actual dose distribution $>D$ for the $i^{th}$ case, $B_i(D)$ = set of all voxels in predicted dose distribution $> D$ for the $i^{th}$ case, and the average $\overline{DSC}_{C,D} = \sum_{i \in c} DSC_{i,D} / Nc$ is computed for isodose levels between 25%- 125% at 10% intervals (to align with $\overline{\delta D}_{j,C}$ and $\sigma_{j,C}$ percentage deciles) and averaged across the $N_C$ plans in the respective cohort C.

To add clinical context to the dose estimation performance, we also utilized common quality metrics employed in brachytherapy plan evaluation[6]. For OARs bladder, rectum, and sigmoid, we used $D_{2cc}$ (dose to the hottest 2cc volume) and for HRCTV we used $D_{90\%}$ (dose to the hottest 90% volume). Model performance on these parameters was quantified using the difference $\Delta D_x = D_{x,actual} - D_{x,predicted}$. Across each cohort C, model prediction of $D_x$ was evaluated with Pearson correlation coefficient $R$, $\overline{\Delta D_x} = \sum_{i \in c} D_{x,i} / N_c$ and the standard deviation of the $\Delta D_{x,C}$ distribution ($\sigma_{\Delta D}$). All $D_x$ were reported in absolute dose (Gy) as this is more commonly used to evaluate OARs during treatment planning. Good performance on discrete quality metric prediction can be seen by $|\overline{\Delta D_x}| \ll \sigma_{\Delta D} \ll D_{Rx}$ where the nominal prescription dose $D_{Rx}$ is 6 Gy (used in 62% of available cases; full range of prescription doses was 5.5-8 Gy across the cohorts).

The out-of-sample test set was used as the primary comparisons to the training set since it was fully independent of the training process; however, the model performance on the validation cases will be shown for completeness.

## Results

### Model Training

In total, the model took 75 hours to train on two GPU nodes for 100 epochs. Training and validation loss were calculated at each epoch. Using the epoch loss data, we selected epoch 80 for the final model because the training loss was still decreasing but the validation error was not **Figure 1**. Predictions on an individual dataset took ~1 minute using one GPU.

### 3D Dose Prediction

Figure 2 shows a randomly selected patient from the test set as an example of the dose estimation performance. The outline of the input contours shows the dose distribution inside the OARs in the axial, sagittal, and coronal planes. To visualize the voxel-by-voxel performance, $\delta D_{xyz}$ is shown for the same viewing planes with the 100% prescription isodose of the actual plan shown as a dashed line. As expected, the large dose differences are confined near the applicator while the agreement outside the 100% isodose surface is good.

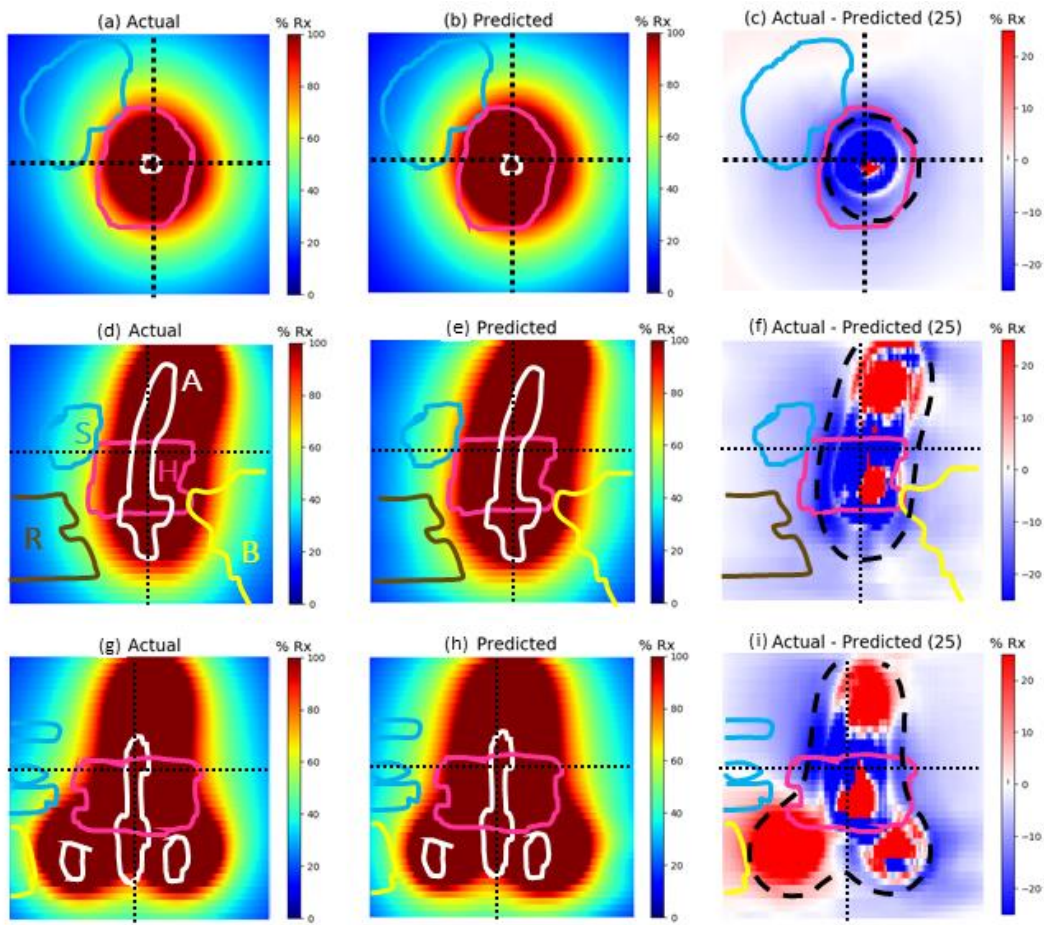

**Fig. 2.** *Axial (top row), sagittal (middle row) and coronal (bottom row) slices show the actual, predicted and dose differences of a randomly chosen plan from the test set. Solid lines indicate contour outlines (R=rectum, B=bladder, S=sigmoid, H=HRCTV, A=applicator), the dotted line represents the actual 100% isodose line.*

In **Figure 3,** average dose differences $\overline{\delta D}_{j,C}[D_l, D_u]$ and standard deviation $\sigma_{j,C}[D_l, D_u]$ are plotted for each structure in the training, validation, and test set. The all voxel error plots in **Figure 3(a-c)** show minimal bias with standard deviation increasing as dose increases. The training set from 20-130% had a mean $\overline{\delta D}$ ranging from -0.3% to +1.0% and a standard deviation increasing from 2% at 20% through 12% at 130%. Across the 20-130% range, the test set bias was more variable than the training set, spanning -0.1% to +4.0%, representing slightly lowered voxel doses in the model-predicted distributions. The test set standard deviation also increased more quickly (4% to 26%). The table of Dice coefficients for isodoses 25% through 125% at 10% intervals was shown in **Figure 3(a-c)** under the corresponding isodose deciles. The isodose dice coefficient is highest at 25% (training: 0.96, test: 0.94) and decreases slightly with dose until 125% (training: 0.91, test: 0.87).

Within the HRCTV, the training cohort voxels error plots have shown mean error $\overline{\delta D}_{j,C}$ spanning from -1.7% to -3.5% across the 40% - 130% range **Figure 3(d-f)**, representing the predicted doses inside the HRCTV to be higher on average (in contrast to the same analysis across all voxels). Also, in contrast to all voxels analysis, the prediction error was lower inside the HRCTV, with $\sigma$ ranging from 5% to 13% across the 40-130% span. The test set had a slightly higher bias ranging from -2.6% to -3.4% with a $\sigma$ of 5% to 19% across 50-130% span (insufficient test set cases were available <50% prescription isodose). In all cohorts there were proportionally fewer cases with HRCTV minimum doses at low isodoses contributing to the voxel error plots.

Inside the bladder **Figure 3(g-i)**, the training cohort voxels error plots show mean bias $\overline{\delta D}$ spanning from -0.7% to 3% across the 20% - 130% range, representing the network-predicted doses inside the bladder to be lower than the actual (in contrast to the same analysis across

HRCTV voxels). Also, compared to the analysis across all voxels, the prediction error was lower inside the bladder, with $\sigma$ ranging from 2.4% to 13% across the 20-130% span. The test set had a slightly higher bias ranging from -2.5% to 0.8% with a $\sigma$ of 3.5% to 13% across 20-130% span.

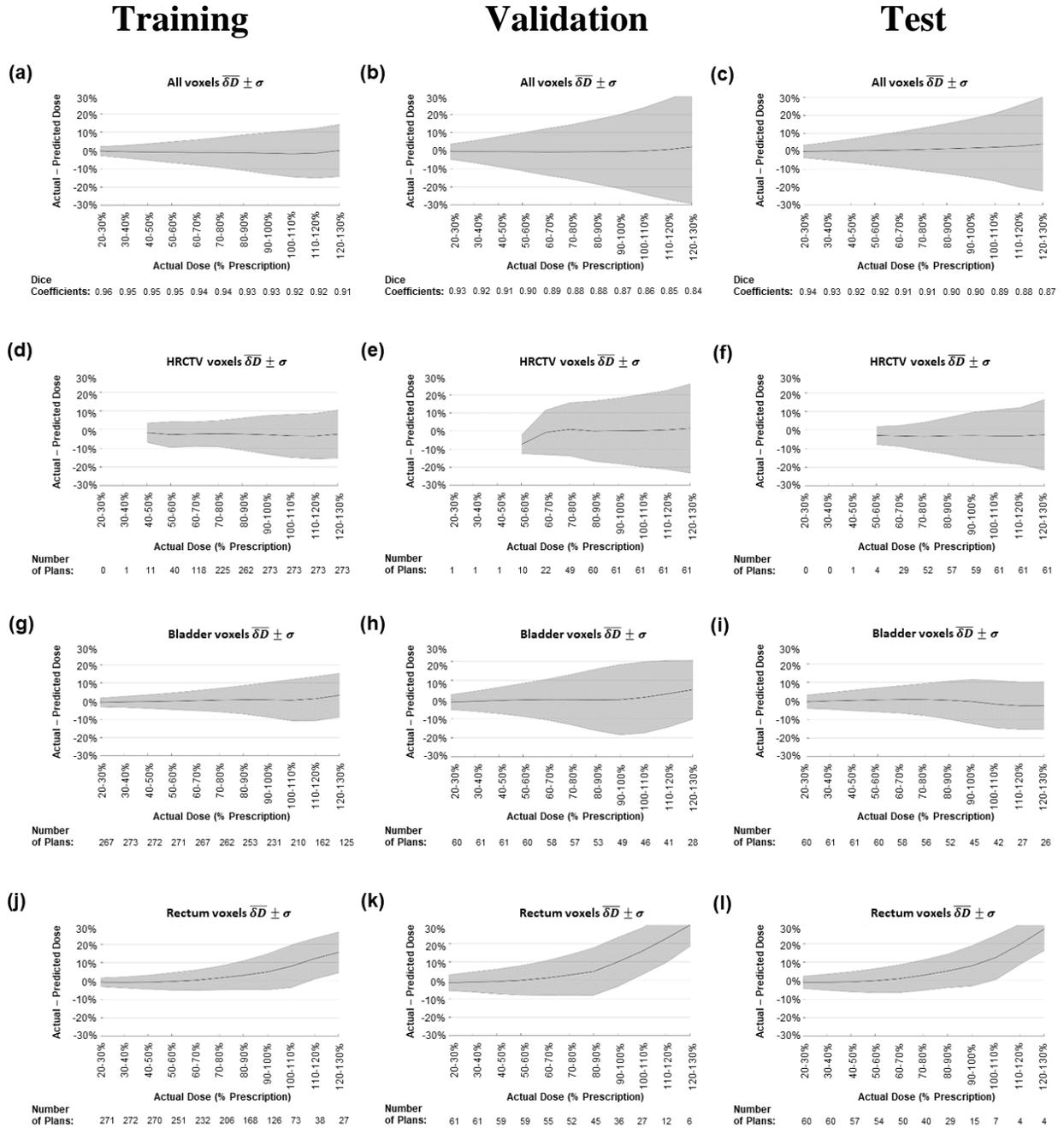

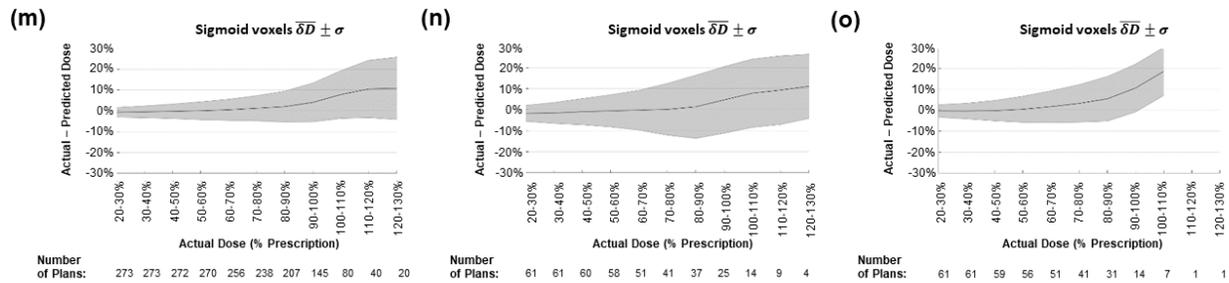

*Fig 3. Average model bias $\overline{\delta D_{j,C}}$ and error $\sigma_{j,C}$ are shown across all voxels (a-c), HRCTV (**d-f**), and OARs (**g-o**) for each cohort. Solid black line denotes $\overline{\delta D_{j,C}}$ and grey error bands denote standard deviation $\sigma_{j,C}$. Dice similarity coefficients for the respective isodose range are shown below the all voxels plot (**a-c**). The number of plans containing voxels in the defined dose regions shown below each structure-specific plot (**d-o**).*

The bias and error for rectum and sigmoid are shown for all cohorts in **Figure 3(j-l)** and **Figure 3(m-o)**, respectively. Both the rectum and sigmoid displayed similar trends of low bias for 20%-70% isodoses, ranging from -0.7%-0.5% and 2.4%-5.5%, respectively. Beyond 70% prescription isodose, the rectum and sigmoid $\overline{\delta D}$ both began increasing steadily with increased dose across all cohorts. Markedly, the bias begins to increase precisely when the number of plans with occupation in these isodose ranges begins to decrease. Within the rectum, the training cohort voxels error plots show mean error $\overline{\delta D}$ increasing from 1.6% to 15.5% across the 70% - 130% range with $\sigma$ ranging from 6.4% to 11% across the 70-130% span. Across the same range, the test set had a higher bias increasing from 2.9% to 28% with an $\sigma$ of 8.2% to 11.5%. As in the case of the bladder, positive mean bias $\overline{\delta D}$ represents the predicted doses inside the rectum and sigmoid to be lower than the actual doses.

Clinical metric prediction exhibited the expected behavior of $|\overline{\Delta D_x}| \ll \sigma_{\Delta D} \ll D_{Rx}$ across all in the training and test cohorts **Figure 4**. The HRCTV $D_{90}$ Pearson correlation coefficients, R, for the training and test set were 0.83 and 0.71, respectively. Corresponding $\overline{\Delta D}_{90} \pm \sigma_{\Delta D}$ were -0.19±0.55 Gy for the training set and -0.09±0.67 Gy for the test set. The slight negative bias across implies higher predicted $D_{90}$ coverage than observed in the actual plans.

Model-predicted $D_{2cc}$s for the OARs were substantially better than HRCTV $D_{90}$ predictions across all OARs and all cohorts. OAR $D_{2cc}$ Pearson correlation coefficients for the bladder, rectum, and sigmoid were 0.91, 0.94, and 0.93 in the training set and 0.88, 0.90, and 0.88 in the test set. $\overline{\Delta D}_{2cc} \pm \sigma_{\Delta D}$ for the bladder were -0.06±0.54 Gy in the training set and -0.17±0.67 Gy in the test set. $\overline{\Delta D}_{2cc} \pm \sigma_{\Delta D}$ for the rectum were -0.03±0.36 Gy in the training set and -0.04±0.46 Gy in the test set. $\overline{\Delta D}_{2cc} \pm \sigma_{\Delta D}$ for the sigmoid were -0.01±0.34 Gy in the training set and 0.00±0.44 Gy in the test set.

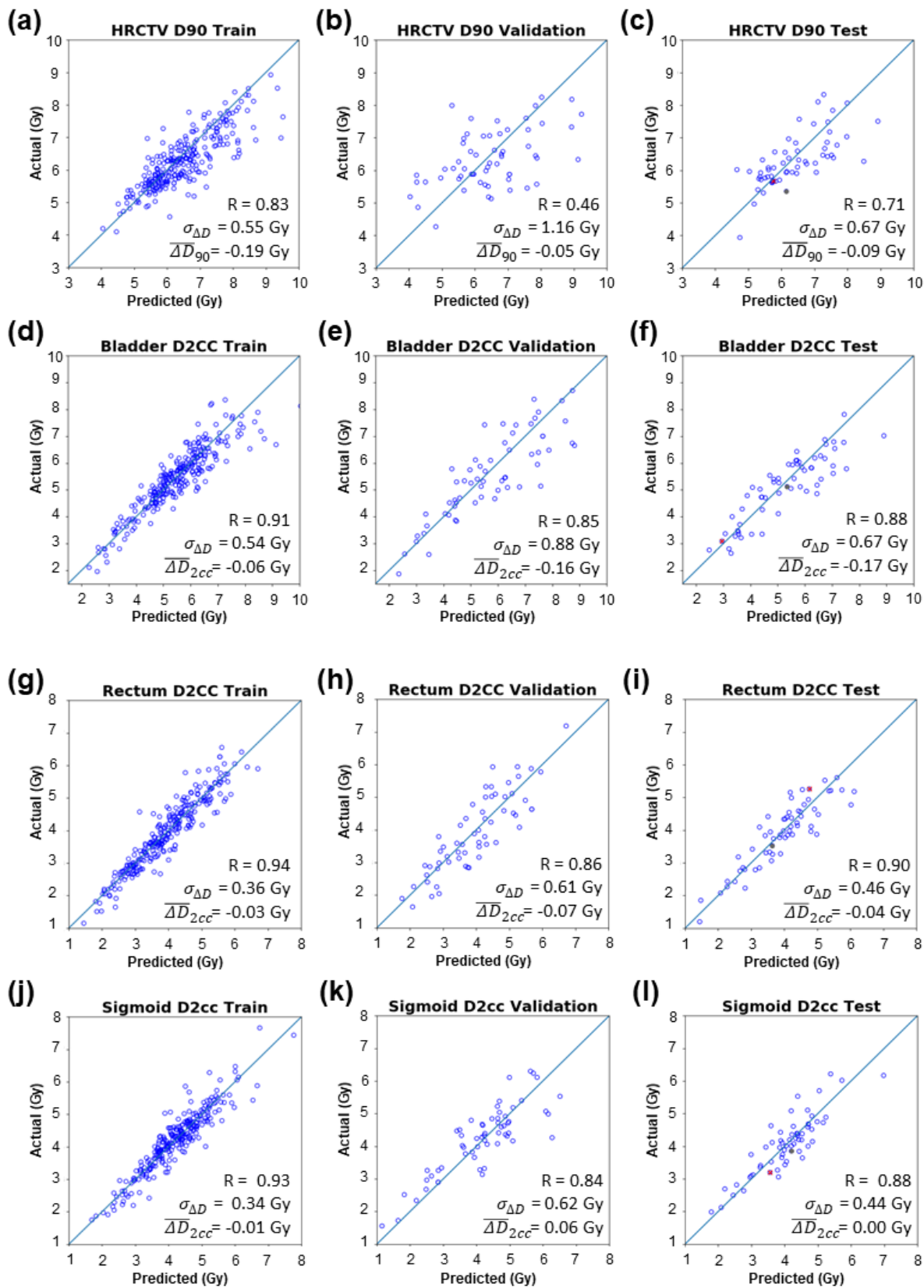

*Fig. 4. Actual vs predicted clinical dose metrics for training, test, and validation datasets. These plots include Pearson correlation coefficient (R), standard deviation ($\sigma_{\Delta D}$) and mean of the clinical metrics ($\overline{\Delta D}$). Blue lines indicate perfect model predictions. Closed dots on test set plots (third column) indicate the test plan that is displayed in Fig 2. Red x indicated plan mentioned in discussion.*

## Discussion

In an effort to standardize cervical cancer brachytherapy, we demonstrate the first knowledge-based 3D dose estimation for cervical cancer brachytherapy using spatial information of OARs, the HRCTV, and applicators to create an accurate voxel-level prediction model. While focused on the widely-used T&O applicator, this method could likely be expandable to other applicators and interstitial situations. This prediction system operates without need for source position digitization. However, if desired this could be easily incorporated by replacing the applicator channel with digitized source positions and retraining the network.

While 3D dose prediction in external beam radiotherapy has been demonstrated across multiple disease sites, the comparatively steep dose gradients and highly inhomogeneous dose distributions in brachytherapy make comparing these modalities difficult. The closest external beam radiotherapy analog would likely be stereotactic radiosurgery (SRS), where Shiraishi *et al*[18] reported voxel prediction error of ~10%, similar to this model's performance shown in Figure 3.

As direct voxel error includes no spatial uncertainty terms, it represents the most stringent evaluation criteria. Prior 3D dose prediction studies have directly employed spatial uncertainty terms such as gamma[28] or isodose dice similarity coefficients in the context of prostate radiotherapy[20,27]. McIntosh and Purdie[20] reported the average Dice coefficient to be 0.88 with a range of 0.82 to 0.93 while Ngyuen *et al*[27] reported the average DSC for their test and cross-validation set as 0.91 and 0.95, respectively. Our model performs similarly with a range of 0.84 to 0.96 for all datasets, with the aforementioned caveat that external beam and brachytherapy dose distributions are not necessarily directly comparable due to fundamental differences in their spatial variation and isodose span.

The visualization of the prediction errors across different structures and isodose domains shown in Figure 3 offers interesting insight into the performance of the system. Noticeably, the rectum and sigmoid structures appear to pick up substantial bias at isodoses near the prescription dose, steadily increasing as the dose increases past 100%. Examining the most extreme cases in this group, one plan in the test set had a mean error of 32% for the rectum voxels with 120-130% prescription dose. Large voxel bias at high doses for this plan can be explained by proximity of the rectum to the ovoids and the very sharp dose gradients in that region. Notably, the HRCTV $D_{90}$, bladder and sigmoid $D_{2cc}$ were within their respective $\sigma_{\Delta D}$ and the rectum $\Delta D_{2cc}$ of 0.50 Gy was only just outside $\sigma_{\Delta D}$ of 0.46 Gy for this cohort, as shown in **Figure 4 (c,f,i,l).** From a clinical perspective, the $D_{2cc}$ and $D_{90}$ predictions are not as biased, indicating that if we act on clinical data any voxel-wise bias at high dose is not as problematic. In addition, DSCs for this plan ranged from [0.88, 0.96] which is better than the test cohort's overall average. This implyies that despite the error in the rectum and sigmoid, isodose similarity for this plan was overall better than the average plan from the test cohort.

The non-zero bias for low occupation, higher isodoses in the OAR structures – even in the training cohort – can be understood from the fact that all cases in the training set contribute to the dose prediction model at all isodoses but only the hottest plans have OAR voxel occupations at near prescription isodoses. There are two explanations for diminishing occupation at higher isodose lines, one purely geometric (some cases have close OAR voxels but all have distant voxels) and one dosimetric (OAR voxel occupation in higher isodose lines biased towards cases where planning did not fully pull isodoses away from these structures, potentially a sign of sub-optimal planning). Both could be at work here, though the dosimetric explanation would result in an increasing bias at high isodoses because while all cases would be contributing to the dose prediction, the remaining sample at greater isodose occupation is biased towards cases where planning did not carve away the dose sufficiently. **Figures 3(g-l)** are suggestive of this interpretation but a definitive demonstration of this would require a full replanning study across a sufficiently large sample which is the focus of future work.

While knowledge-based dose estimation is relatively new to the field of brachytherapy and cervical cancer, it is possible to compare our clinical metric prediction to Yusufaly *et al*[17] where a knowledge-based organ-at-risk DVH estimation modeling approach was used on the same patient cohort in this work (though training, validation, and test set splits were different). OAR $D_{2cc}$ predictions from the 3D neural network vs the 1D DVH estimates show similar accuracy: bladder $\sigma_{\Delta D}$ was 0.67 Gy vs. 0.61 Gy, rectum $\sigma_{\Delta D}$ 0.46 Gy vs. 0.47 Gy, and sigmoid $\sigma_{\Delta D}$ 0.44 Gy vs. 0.47 Gy. Obviously, the 3D neural network approach offers more information – HRCTV predictions, voxel-level estimates, visualized distributions – but it is notable that the clinical metric prediction accuracy was essentially unchanged with this approach. This could be explained by unresolved plan quality variations in the sample; replanning of the significant portion of the patient cohort and then subsequent retraining of the model is needed to determine if the lack of improvement in this aspect of the dose predictions is due to plan quality variations or to some other explanation.

There are some existing limitations with our current approach; our data is restricted to one institution and only includes T&O cases. The current approach also requires that the applicator be inserted to make predictions; however, it can operate without the need for source position digitization.

Despite current limitations, the metrics analyzed demonstrate that the model could serve as a useful quality control tool for T&O brachytherapy and, potentially, as the basis for automated knowledge-based planning.

## Conclusion

This study demonstrates the ability to predict accurate 3D dose distributions for T&O brachytherapy using a deep learning framework.

## Acknowledgements

We thank Kelly Kisling, Xenia Ray, and Aaron Babier for helpful discussions. This work was supported by Pedal the Cause and the Agency for Healthcare Research and Quality (AHRQ R01HS025440).

## Conflict of Interest Statement

KLM reports honoraria and personal fees from Varian Medical Systems, outside the submitted work. KLM holds two patents (Developing Predictive Dose-Volume Relationships for a Radiotherapy Treatment and Knowledge-based Prediction of Three-Dimensional Dose Distributions), licensed to Varian Medical Systems. AS reports personal fees from Courage Health, Inc., outside the submitted work. JM reports personal fees from AstraZeneca, grants from NRG Oncology, grants from GOG Foundation, personal fees from Varian Medical Systems, outside the submitted work.